\documentstyle{article}

\newcommand{\be}{\begin{equation}}
\newcommand{\ee}{\end{equation}}
\newcommand{\ba}{\begin{eqnarray}}
\newcommand{\ea}{\end{eqnarray}}
\newcommand{\no}{\nonumber\\}

\newcommand{\grts}{\raise.3ex\hbox{$>$\kern-.75em\lower1ex\hbox{$\sim$}}}
\newcommand{\lets}{\raise.3ex\hbox{$<$\kern-.75em\lower1ex\hbox{$\sim$}}}

\title{A new model for the quark mass matrices}

\author{L.\ Lavoura \\
\small Universidade T\'ecnica de Lisboa \\
\small Centro de F\'\i sica das Interac\c c\~oes Fundamentais \\
\small Instituto Superior T\'ecnico, 1049-001 Lisboa, Portugal}

\date{28 July 1999}

\begin{document}
\maketitle

\begin{abstract}
I present a new model for the quark mass matrices,
which uses four scalar doublets
together with a horizontal symmetry $S_3 \times Z_3$.
The model is inspired on a suggestion made a few years ago by Ma,
but it is different.
The predictions $\left| V_{ts} \right| \approx m_s / m_b$
and $\left| V_{ub} / V_{cb} \right| > 0.085$ are obtained.
Flavour-changing neutral Yukawa interactions
do not exist in the down-type-quark sector.
\end{abstract}

A few years ago,
Ernest Ma \cite{ma} put forward a model for the quark mass matrices
based on the discrete symmetry $S_3 \times Z_3$.
His model is characterized
by the absence of flavour-changing neutral Yukawa interactions (FCNYI)
from the charge-$2/3$-quark sector.
Ma's model predicts rather low values
for both $\left| V_{ub} / V_{cb} \right|$ and $m_s / m_d$;
those values are moreover correlated,
with a larger $\left| V_{ub} / V_{cb} \right|$
implying a lower $m_s / m_d$,
and vice-versa \cite{lavoura1}.
In spite of this problem,
Ma's mass matrices should be praised:
they are not postulated as an {\it Ansatz},
a ``scheme'',
or a ``texture'',
rather they follow from a complete model,
with a well-defined field content and a well-defined internal symmetry.
As a consequence,
Ma's model---which needs to be supplemented by soft symmetry breaking
in the scalar potential \cite{lavoura2}---is self-contained
and consistent from the point of view of quantum field theory.
This situation contrasts with the one typical of many,
sometimes otherwise quite successful,
schemes or textures that have been proposed for the quark mass matrices.
Most of those schemes cannot be justified
in terms of a complete theory \cite{kitchen sink}.
In this very important respect,
Ma's model is clearly superior.

In this Brief Report I remark that the role of the up-type
and down-type quarks in Ma's model may be interchanged,
one then obtaining another viable model for the mass matrices.
The new model enjoys the same features
of self-containedness and consistency as Ma's one.
It leads to quite distinct predictions for the quark masses and mixings.
A very clear-cut prediction is $\left| V_{ts} \right| \approx m_s /m_b$;
as a consequence of this relation and of the measured value
of $\left| V_{cb} \right| \approx \left| V_{ts} \right|$,
the strange-quark mass must be in the upper part of its allowed range.
It also predicts $\left| V_{ub} / V_{cb} \right| > 0.085$.
FCNYI are absent from the charge-$- 1/3$-quark sector,
impeding tree-level contributions to the mass differences
in the $K^0$--$\overline{K^0}$ and $B_d^0$--$\overline{B_d^0}$ systems,
and to the CP-violating parameter $\epsilon$.

In my model there are four Higgs doublets $\phi_a$ ($a = 1, 2, 3, 4$)
and two $Z_3$ symmetries.
I denote $p_{Ri}$  ($i = 1, 2, 3$) the right-handed charge-$2/3$ quarks,
$n_{Ri}$ the right-handed charge-$- 1/3$ quarks,
and $q_{Li} = \left( p_{Li}, n_{Li} \right)^T$
the doublets of left-handed quarks.
The quantum numbers of the various fields
under $Z_3^{(1)}$ and $Z_3^{(2)}$ are given in Table~\ref{quantum numbers}.
\begin{table}
{\caption{Quantum numbers of the fields under the two $Z_3$ symmetries,
with $\omega \equiv \exp \left( 2 i \pi / 3 \right).$}
\label{quantum numbers}}
\begin{center}
{\begin{tabular}{|c|ccccccccccccc|}
\hline
 \hfil & $\phi_1$ & $\phi_2$ & $\phi_3$ & $\phi_4$ &
$q_{L1}$ & $q_{L2}$ & $q_{L3}$ &
$p_{R1}$ & $p_{R2}$ & $p_{R3}$ &
$n_{R1}$ & $n_{R2}$ & $n_{R3}$ \\
\hline
$Z_3^{(1)}$ & $\omega^2$ & $\omega$ & $1$ & $1$ &
$1$ & $\omega^2$ & $\omega$ &
$1$ & $\omega$ & $\omega^2$ &
$1$ & $\omega$ & $\omega^2$ \\
$Z_3^{(2)}$ & $1$ & $1$ & $\omega$ & $\omega^2$ &
$\omega^2$ & $1$ & $1$ &
$1$ & $\omega^2$ & $\omega^2$ &
$\omega$ & $1$ & $1$ \\
\hline
\end{tabular}}
\end{center}
\end{table}

Besides $Z_3^{(1)}$ and $Z_3^{(2)}$,
there is one further horizontal symmetry,
which effects the interchanges
\be
\phi_1 \leftrightarrow \phi_2, \
q_{L2} \leftrightarrow q_{L3}, \
p_{R2} \leftrightarrow p_{R3}, \
n_{R2} \leftrightarrow n_{R3},
\label{Z2}
\ee
and leaves all other fields invariant.
This symmetry commutes with $Z_3^{(2)}$
but it does not commute with $Z_3^{(1)}$.
Hence,
the internal-symmetry group of the model is $S_3 \times Z_3^{(2)}$.

As a consequence of this internal symmetry,
the quark mass matrices take the form
\be
M_p = \left( \begin{array}{ccc}
y_1 v_3^\ast & y_2 v_2^\ast & y_2 v_1^\ast \\
y_3 v_2^\ast & 0 & y_4 v_4^\ast \\
y_3 v_1^\ast & y_4 v_4^\ast & 0
\end{array} \right), \
M_n = \left( \begin{array}{ccc}
y_5 v_3 & 0 & 0 \\ 0 & y_6 v_2 & 0 \\ 0 & 0 & y_6 v_1
\end{array} \right),
\label{mass matrices 1}
\ee
where the Yukawa coupling constants $y_{1\mbox{--}6}$
and the vacuum expectation values (VEVs)
$v_a = \left< 0 \left| \phi_a^0 \right| 0 \right>
= \left| v_a \right| \exp \left( i \theta_a \right)$
are in general complex.
One identifies $\left| y_5 v_3 \right| = m_d$,
the mass of the down quark,
while $\left| y_6 v_2 \right| = m_s$ and $\left| y_6 v_1 \right| = m_b$
are the masses of the strange quark and of the bottom quark,
respectively.
Thus,
$\left| v_2 / v_1 \right| = m_s / m_b \equiv r$.
This ratio of VEVs being different from $1$,
the internal symmetry of Eq.~(\ref{Z2})
is spontaneously broken.\footnote{The spontaneous breaking
of the interchange symmetry of Eq.~(\ref{Z2})
follows from its soft breaking in the scalar potential \cite{lavoura2},
which is achieved through the introduction of a term
$\mu \left( \phi_1^\dagger \phi_1 - \phi_2^\dagger \phi_2 \right)$.}

One may eliminate most of the phases in the mass matrices
by means of rephasings of the quark fields,
obtaining
\be
M_p = \left( \begin{array}{ccc}
f & r g e^{i \psi} & g \\ r h e^{i \psi} & 0 & a \\ h & a & 0
\end{array} \right), \
M_n = \left( \begin{array}{ccc}
m_d & 0 & 0 \\ 0 & m_s & 0 \\ 0 & 0 & m_b
\end{array} \right),
\label{mass matrices 2}
\ee
where $a$,
$f$,
$g$,
and $h$ are real and non-negative.
$M_p$ is bi-diagonalized by the Cabibbo--Kobayashi--Maskawa matrix $V$
and another unitary matrix,
$U_R^p$:
\be
V M_p U_R^p = \mbox{diag} \left( m_u, m_c, m_t \right).
\label{bi-adiagonalization}
\ee
Thus,
\be
H \equiv M_p M_p^\dagger =
\left( \begin{array}{ccc}
f^2 + g^2 \left( 1 + r^2 \right) & a g + r f h e^{- i \psi} &
f h + r a g e^{i \psi} \\ a g + r f h e^{i \psi} & a^2 + r^2 h^2 &
r h^2 e^{i \psi} \\ f h + r a g e^{- i \psi} & r h^2 e^{- i \psi} & a^2 + h^2
\end{array} \right) = V^\dagger \left( \begin{array}{ccc}
m_u^2 & 0 & 0 \\ 0 & m_c^2 & 0 \\ 0 & 0 & m_t^2 \end{array} \right) V,
\label{Hp}
\ee
and one immediately sees that
\be
\frac{H_{33} - H_{22}}
{\left| H_{23} \right|} = \frac{1}{r} - r
= \frac{m_b}{m_s} - \frac{m_s}{m_b}.
\label{exact relation}
\ee
Using $H_{22} \ll H_{33} \approx m_t^2$
and $ \left| H_{23} \right| \approx m_t^2 \left| V_{ts} \right|$,
one finds the main prediction of this model,
\be
\left| V_{ts} \right| \approx \frac{m_b m_s}{m_b^2 - m_s^2}
\approx \frac{m_s}{m_b}.
\label{approximate relation}
\ee
Equation (\ref{approximate relation}) is almost exact.
In practice,
we may write it with $\left| V_{ts} \right|$
substituted by the more interesting parameter $\left| V_{cb} \right|$,
obtaining the slightly worse approximation
\be
\left| V_{cb} \right| \approx \frac{m_s}{m_b}.
\label{approximate relation 2}
\ee

I use the quark masses renormalized at $1\, \mbox{GeV}$ \cite{gasser,leutwyler}
\be
m_s = \left( 175 \pm 25 \right) {\rm MeV}, \
m_b = \left( 5.3 \pm 0.1 \right) {\rm GeV}.
\label{ms e mb}
\ee
The scale uncertainty on the light-quark masses is substantial,
while their ratios are relatively well known;
in particular \cite{leutwyler},
\be
\frac{m_s}{m_u} = 34.4 \pm 3.7.
\label{razao de ms e mu}
\ee
Comparing Eqs.~(\ref{approximate relation 2})
and (\ref{ms e mb}) with the experimental value \cite{PDG}
\be
\left| V_{cb} \right| = 0.0395 \pm 0.0017,
\label{Vcb}
\ee
one sees that the main prediction of the model is quite well verified;
as a matter of fact,
the smallness of the error bar in Eq.~(\ref{Vcb}) allows us
to constrain $m_s$ to be in the highest part of its allowed range:
\be
m_s \left( 1\, {\rm GeV } \right) \grts\ 190\, {\rm MeV}.
\label{ms}
\ee

In order to find out other predictions of the model
one must treat it numerically.
One easily concludes that $h \approx m_t$
and the phase $\psi$ is very close to zero.
Contrary to what happens in most {\it Ans\"atze} and textures,
$\left| V_{us} \right|$ is not related to quark-mass ratios,
rather it must be fitted to its experimental value $0.22$.
The exact value of the top-quark mass $m_t$,
being quite high,
is practically irrelevant.
One finds
\be
\left| \frac{V_{ub}}{V_{cb}} \right| > 0.085,
\label{razao de Vub e Vcb: previsao}
\ee
to be compared with the experimental value \cite{PDG}
\be
\left| \frac{V_{ub}}{V_{cb}} \right| = 0.08 \pm 0.02.
\label{razao de Vub e Vcb}
\ee
My model easily accomodates $\left| V_{ub} / V_{cb} \right|$
as large as $0.3$.\footnote{The error bar
in Eq.~(\ref{razao de Vub e Vcb}) is probably under-estimated \cite{pene},
as there is a substantial uncertainty
in the theoretical modelling of $b \rightarrow u$ decays.
Maybe one should not exclude the possibility
that $\left| V_{ub} / V_{cb} \right|$ is substantially higher than $0.1$.}
On the other hand,
$\left| V_{ub} / V_{cb} \right| \lets\, 0.09$ is only marginally possible.

The CP-violating invariant $J$ \cite{jarlskog}
is small because $\psi$ is so close to zero.
One obtains $\left| J \right| < 3.5 \times 10^{-5}$,
but usually $J$ is barely enough
to account for the observed value of $\epsilon$.
This is not a problem,
since there are in the model extra CP-violating box diagrams,
in particular those with virtual charged scalars.

In order to work out the Yukawa interactions,
one should first expand the scalar doublets as
\be
\phi_a = e^{i \theta_a} \left( \begin{array}{c} \phi_a^+ \\
\left| v_a \right| + {\displaystyle
\frac{ \rho_a + i \eta_a}{\sqrt{2}}} \end{array}
\right),
\label{doublets}
\ee
where $\rho_a$ and $\eta_a$ are Hermitian fields.
Their Yukawa interactions are given by
\ba
{\cal L}_{\rm Y}^{\rm (q)} &=& \cdots
- \frac{\left( \rho_1 + i \eta_1 \right) m_b \overline{b_L} b_R}
{\sqrt{2} \left| v_1 \right|}
- \frac{\left( \rho_2 + i \eta_2 \right) m_s \overline{s_L} s_R}
{\sqrt{2} \left| v_2 \right|}
- \frac{\left( \rho_3 + i \eta_3 \right) m_d \overline{d_L} d_R}
{\sqrt{2} \left| v_3 \right|}
\no                  & &
- \left( \overline{u_L},\ \overline{c_L},\ \overline{t_L} \right) V
\left( \begin{array}{ccc}
{\displaystyle \frac{\rho_3 - i \eta_3}
{\sqrt{2} \left| v_3 \right|}} d &
{\displaystyle \frac{\rho_2 - i \eta_2}
{\sqrt{2} \left| v_2 \right|}} r g e^{i \psi} &
{\displaystyle \frac{\rho_1 - i \eta_1}
{\sqrt{2} \left| v_1 \right|}} g \\*[3mm]
{\displaystyle \frac{\rho_2 - i \eta_2}
{\sqrt{2} \left| v_2 \right|}} r b e^{i \psi} &
0 &
{\displaystyle \frac{\rho_4 - i \eta_4}
{\sqrt{2} \left| v_4 \right|}} a \\*[3mm]
{\displaystyle \frac{\rho_1 - i \eta_1}
{\sqrt{2} \left| v_1 \right|}} b &
{\displaystyle \frac{\rho_4 - i \eta_4}
{\sqrt{2} \left| v_4 \right|}} a &
0
\end{array} \right)
U_R^p \left( \begin{array}{c} u_R \\ c_R \\ t_R \end{array} \right).
\label{Yukawa interactions}
\ea
In the first line of Eq.~(\ref{Yukawa interactions}) one observes
the absence of FCNYI with the down-type quarks.
Unfortunately,
the physical neutral scalars result from
an unspecified orthogonal rotation of the $\rho_a$ and $\eta_a$---wherein
$\sum_{a=1}^4 \left| v_a \right| \eta_a$ is a Goldstone boson.
This is equivalent to saying that the symmetries of the model
do not determine the masses and mixings of the neutral scalars.
Moreover,
the constraints $\left| v_2 / v_1 \right| = m_s / m_b$
and $\sum_{a=1}^4 \left| v_a \right|^2 = \left( 2 \sqrt{2} G_F\right)^{-1}$
are insufficient to determine the four $\left| v_a \right|$.
Under these conditions,
any attempt at an evaluation of the effects
of the neutral Yukawa interactions---in particular
enhanced $D^0$--$\overline{D^0}$ mixing,
which might be an interesting consequence
of the present model---cannot be rigorous
and has little genuine value.\footnote{The studies
\cite{pal,lavoura2} of the FCNYI
in Ma's model suffer from the same limitation:
many assumptions about the values of the $\left| v_a \right|$
and of the neutral-scalar mixings had to be done.}
The same may of course be said about an evaluation
of the effects of the charged Yukawa interactions,
including their contribution to $\epsilon$.

In conclusion,
I have shown that,
in Ma's model for the quark mass matrices,
the roles of the up-type and down-type quarks may be interchanged,
one then obtaining a different viable model.
The new model makes the sharp prediction
$\left| V_{cb} \right| \approx m_s / m_b$
and forces $m_s$ and $\left| V_{ub} / V_{cb} \right|$
to be close to the upper end of their allowed ranges.
Flavour-changing neutral Yukawa interactions
are absent from the charge-$-1/3$-quark sector.
The model has the distinctive advantages
of having a well-defined field content and horizontal symmetry,
and of not containing any poorly justified assumptions.

\end{document}